\documentclass[conference]{IEEEtran}
\IEEEoverridecommandlockouts
\usepackage{epsf}
\usepackage{nopageno}
\usepackage{graphicx}
\usepackage{epsfig,latexsym,amsmath,epsf,amssymb,amsfonts}
\usepackage{amssymb}
\usepackage{caption}
\usepackage{epsfig}
\usepackage{epstopdf}
\usepackage{color}
\usepackage[utf8]{inputenc}
\usepackage{amsmath}
\usepackage{epstopdf}
\usepackage{cite}
\usepackage[T1]{fontenc}
\usepackage[applemac]{inputenx}
\usepackage[bookmarks=false]{hyperref}
\usepackage{soul}

\usepackage{mathptmx}
\usepackage{setspace}
\usepackage{subfig}
\usepackage{subcaption}
\usepackage{multicol}%
\usepackage{enumerate}
\usepackage{lettrine}
\usepackage[inkscapelatex=false]{svg}
\usepackage[ruled]{algorithm2e}
\usepackage{algpseudocode}

\DeclareMathAlphabet{\mathcal}{OMS}{cmsy}{m}{n}
\SetMathAlphabet{\mathcal}{bold}{OMS}{cmsy}{b}{n}

\begin{document}
\title{\Huge{Towards SAFE-ISAC: STAR-RIS-Aided Joint Jamming Suppression and Target Concealment}}

\vspace{-0.3cm}
\author{\large{Radwa Sultan}\\
\begin{tabular}{c}
\small{Electrical and Computer Engineering Department, Kennesaw State University, Marietta, Georgia} \\

\end{tabular}}

\maketitle
\thispagestyle{empty}

	\begin{abstract}

  Designing robust architectures that can mitigate sophisticated attacks is now a key priority for modern wireless systems. This paper investigates a single-cell bistatic integrated sensing and communication (ISAC) network facing simultaneous coordinated active jamming and malicious detection. These threats aim to disrupt the downlink communication and detect the presence of the ISAC target, respectively. To counter these attacks, we propose the SAFE-ISAC framework, which utilizes a simultaneous transmit and reflect reconfigurable intelligent surface (STAR-RIS) to jointly suppress jamming power and reduce the malicious detector's Signalto-Interference-plus-Noise Ratio (SINR). We formulate a joint minimization problem for jamming gain and detection probability by optimizing the STAR-RIS reflection and transmission responses. This non-convex problem is decoupled into two subproblems: i) malicious detection mitigation in the transmission subspace, solved using the Dinkelbach method and Semidefinite Programming (SDP) relaxation, and ii) jamming suppression in the reflection subspace, addressed via Polak-Reib\'{e}re Riemannian conjugate gradient algorithm. Numerical results validate that the proposed scheme effectively achieves jamming mitigation and target concealment while meeting all communication and sensing Quality-of-Service (QoS) requirements, compared to existing benchmarks.
  
	\end{abstract}


	%
%
\section{Introduction}\label{tab:intro}

The rapid growth in the number of users, devices, and services in next-generation cellular networks calls for higher spectrum efficiency \cite{6G}. To address this challenge, several frameworks have been proposed, including non-orthogonal multiple access (NOMA) techniques \cite{noma1,noma2}, orbital angular momentum (OAM) \cite{oam1,oam2}, full-duplex (FD) communication \cite{fd1}, and reconfigurable intelligent surfaces (RISs) \cite{fd2,fd3}. In addition, the integration of sensing and communication, referred to as dual-functional networks, has recently attracted significant attention as it can mitigate spectrum congestion between sensing and communication functionalities and thereby improve overall spectrum efficiency \cite{isac1,isac2}. However, this integration introduces several challenges, including waveform design, interference management, and security \cite{isac3}.

In particular, enabling ISAC in RIS-aided networks is a promising approach, as the channel can be dynamically controlled through RIS, leading to improved sensing and communication performance. In \cite{ris1}, to optimize the radar signal-to-interference-plus-noise ratio (SINR), the authors proposed a new multifunctional RIS. By jointly optimizing the base station (BS) transmit beamforming and RIS configurations under multiple power and Quality-of-Service (QoS) constraints, the authors demonstrated that RIS performance varies across operational protocols, where the energy-splitting (ES) protocol offers the best performance.  In \cite{ris2}, the authors considered physical layer security in an RIS-aided ISAC system. The framework is optimized to maximize the minimum sensing performance subject to power and secrecy rate constraints. In \cite{ris3}, a joint antenna selection and beamforming design framework is proposed. By adopting a cuckoo search-based scheme, the antennas with the highest channel gains are selected. Afterward, the weighted sum rate is maximized subject to the power constraints of the radar, the communication transmitter, and the RIS. Furthermore, significant research focus has been directed toward security and privacy in RIS-aided ISAC systems. In \cite{refHide}, the authors utilized an RIS to ensure the target remains detectable only to legitimate sensors. This is achieved by jointly optimizing the RIS reflection and sensing elements, the receive beamformer, and the BS transmit beamformer, all while maintaining the minimum required sensing performance at the legitimate sensors. In \cite{sec1}, a covert transmission problem is studied. The authors adopted a Twin Delayed Deep Deterministic Policy Gradient (TD3) algorithm to joint optimize the covert beamforming and the RIS phase shifts. In \cite{sec2}, an active simultaneously transmitting and reflecting reconfigurable intelligent surface (STAR)-RIS aided ISAC system is considered. The authors assumed that the targets are attackers. The first attacker is a warden who potentially detects of covert transmission and the second is an eavesdropper who would intercept the information of secure users. The authors proposed a joint BS and RIS beamforming optimization to maximize the minimum beampattern gain subject to communication security and  covertness constraints.

In this paper, we consider a STAR-RIS-aided bistatic ISAC system under the presence of simultaneous coordinated communication jamming and malicious target detection. On the sensing side, we assume the absence of a direct link between the ISAC transmitter (TX) and the target. Consequently, with the aid of the STAR-RIS, the probing signal is forwarded from the ISAC-TX to the target. Meanwhile, a malicious detector transmits its own probe signal in an attempt to localize the target. On the communication side, the downlink (DL) communication between the ISAC-TX and the communication users is under a jamming attack attempting to disrupt the communication quality. To mitigate this coordinated attack \cite{ref222}, we propose a STAR-RIS-aided joint jamming mitigation and target concealment optimization problem. The main problem is then divided into two subproblems. The first subproblem, in the STAR-RIS transmission subspace, aims to minimize the malicious detector's detection probability under the detection constraints of the legitimate detector. By applying Fractional Programming (FP) and Semidefinite Relaxation (SDR), the subproblem is reformulated into a standard Semidefinite Programming (SDP) problem. The second subproblem, in the STAR-RIS reflecting subspace, aims to minimize the jamming interference power under the communication channel gain constraint. The constraints are handled using a penalty approach and the problem is solved using the Riemannian Polak-Reib\'{e}re conjugate gradient algorithm. Via numerical analysis, we show the effectiveness of the STAR-RIS in mitigating simultaneous jamming and detection attacks when compared to the conventional reflective-only RIS and the random-phase STAR-RIS. 

\indent The rest of this paper is organized as follows. The system model is presented in Section \ref{tab:sysMod}. We present the proposed optimization scheme in Section \ref{tab:probform}. The system is numerically analyzed in Section \ref{tab:numAna}. Finally, the paper is concluded in Section \ref{tab:conc}.

\begin{figure} [t]
    \centering
    \includegraphics[width=10cm,height=9cm]{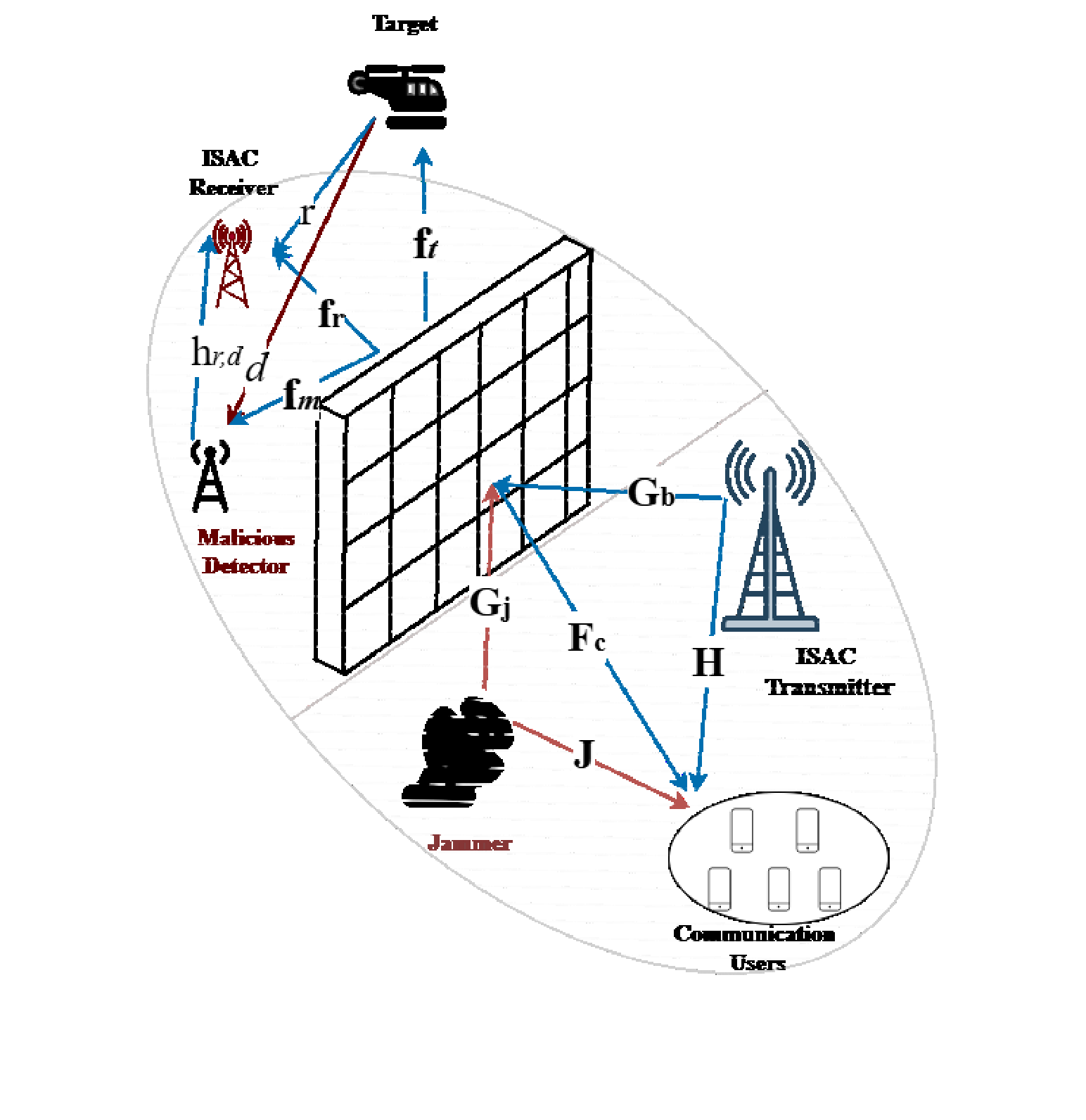} 
    \caption{STAR RIS-aided ISAC System.}
    \label{fig:sysmodel}
  \end{figure}

\section{System Model} \label{tab:sysMod}

In this paper, we investigate a bistatic ISAC system consisting of a multiple-input-multiple-output (MIMO) ISAC BS transmitter with $N$ antennas and a single-antenna ISAC receiver, interconnected via a backhaul link.  The system serves $K$ single-antenna DL users and senses a single target. The network is subject to a coordinated adversarial attack, in which an $N_j$-antenna jammer and a malicious detector cooperate to jointly disrupt communication and sensing operations. Consequently, the network is required to conceal the target from potential malicious detection while simultaneously protecting the communication data against jamming interference. To meet these challenges, the ISAC system utilizes an $L$- element equal ES STAR-RIS to suppress jamming interference and reduce the target's detectability by the malicious detector. The system model is illustrated in Fig. \ref{fig:sysmodel}. 

We assume that the direct path between the ISAC transmitter and the sensing target is blocked, i.e., it experiences deep fading. Accordingly, the sensing signal is transmitted to the target exclusively via the STAR-RIS transmission. The echo signals are subsequently processed at a dedicated ISAC receiver located in the vicinity of the target. Simultaneously, a malicious detector attempts to intercept the echo signals to identify the target. On the other hand, in the STAR-RIS reflection subspace, a jammer operates to disrupt communication between the ISAC transmitter and the network users. It is assumed that the jammer and the malicious detector cooperate such that the jamming signal is nulled at the malicious detector, ensuring that the target's echoes reception remains unaffected at the malicious detector \cite{ref222}.

\subsection{Communication Model}

Under the aforementioned assumptions and given that the communication lies in the STAR-RIS reflection subspace, the effective channels from the ISAC transmitter to the communication users, $\mathbf{H}_{\text{eff}} \in \mathbb{C}^{K\times N}$, and from the jammer to the communication users, $\mathbf{J}_{\text{eff}} \in \mathbb{C}^{K\times N_j}$, are given respectively by

\begin{equation}
    \begin{aligned}
    \mathbf{H}_{\text{eff}} &= \mathbf{H}+\mathbf{F}_c\boldsymbol{\Psi}_r\mathbf{G}_b, \\
     \mathbf{J}_{\text{eff}} &= \mathbf{J}+\mathbf{F}_c\boldsymbol{\Psi}_r\mathbf{G}_j, \\
    \end{aligned}
\end{equation}

\noindent where $\mathbf{H} \in \mathbb{C}^{K\times N}$ is the direct channel between the ISAC transmitter and the network users, $\mathbf{F}_c \in \mathbb{C}^{K \times L}$ is the channel between the STAR-RIS and the users, $\mathbf{G}_b \in \mathbb{C}^{L \times N}$ is the channel between the ISAC transmitter and the STAR-RIS,  $\mathbf{J} \in \mathbb{C}^{K\times N_j}$ is the direct channel between the jammer and the network's users, $\mathbf{G}_j \in \mathbb{C}^{L \times N_j}$ is the channel between the jammer and the STAR-RIS, and $\boldsymbol{\Psi}_r$ is the STAR-RIS reflection $L \times L$ response matrix. The value of $\boldsymbol{\Psi}_r$ is given by 

\begin{equation}
\boldsymbol{\Psi}_r = 
  \begin{bmatrix}
    e^{\mathit{j} \theta_r^1} & & & \\
    & e^{\mathit{j} \theta_r^2} & &  \\
    &  & \ddots & \\
    & &  & e^{\mathit{j} \theta_r^L}
  \end{bmatrix}.
\end{equation}

Accordingly, the achievable sum rate in the presence of the jammer is given by 
\begin{equation}
\label{tab:SINRDL}
R = \log_2 \text{det} \bigg(\mathbf{\it{I}}+p_c\mathbf{H}_{\text{eff}}\mathbf{H}_{\text{eff}}^H\big(\sigma_n^2\mathbf{\it{I}}+p_j\mathbf{J}_{\text{eff}}\mathbf{J}_{\text{eff}}^H\big)^{-1}\bigg),
\end{equation}

\noindent where $p_c$ is the ISAC transmitter power, $p_j$ is the jamming power, and $\sigma_n^2$ is the AWGN noise power.

\subsection{Sensing Model}

As we mentioned, we consider a bistatic ISAC system in which the ISAC transmitter and receiver are spatially separated. Moreover, the direct path between the target and the ISAC transmitter is blocked. The ISAC receiver, the malicious detector, and the target are assumed to reside in the STAR-RIS reflection subspace. Furthermore, the malicious detector transmits its own probing signal to help in target detection. Therefore, the effective channel from the ISAC transmitter to the ISAC receiver, $\mathbf{h}_{\text{R-T|eff}} \in \mathbb{C}^{1\times N}$, the effective channel from the jammer to the ISAC receiver, $\mathbf{h}_{\text{R-J|eff}} \in \mathbb{C}^{1\times N_j}$, from the ISAC transmitter to the target, $\mathbf{h}_{\text{s|eff}} \in \mathbb{C}^{1\times N}$, and from the ISAC transmitter to the malicious detector, $\mathbf{i}_{\text{eff}} \in \mathbb{C}^{1\times N}$, are respectively given by

\begin{equation}
    \begin{aligned}
    &\mathbf{h}_{\text{R-T|eff}} = \mathbf{f}_r\boldsymbol{\Psi}_t\mathbf{G}_b, 
    \mathbf{h}_{\text{R-J|eff}} = \mathbf{f}_r\boldsymbol{\Psi}_t\mathbf{G}_j, \\
    &\mathbf{h}_{\text{s|eff}} = \mathbf{f}_t\boldsymbol{\Psi}_t\mathbf{G}_b, 
     \mathbf{i}_{\text{eff}} = \mathbf{f}_m\boldsymbol{\Psi}_t\mathbf{G}_b, \\
    \end{aligned}
\end{equation}

\noindent where $\mathbf{f}_r \in \mathbb{C}^{1 \times L}$ is the channel vector between the STAR-RIS and the ISAC receiver, $\mathbf{f}_t \in \mathbb{C}^{1 \times L}$ is the channel between the target and the STAR-RIS, $\mathbf{f}_m \in \mathbb{C}^{1 \times L}$ is the channel between the malicious target and the STAR-RIS, and $\boldsymbol{\Psi}_t$ is the STAR-RIS transmission $L \times L$ response  matrix, which is given by 

\begin{equation}
\boldsymbol{\Psi}_t = 
  \begin{bmatrix}
    e^{\mathit{j} \theta_t^1} & & & \\
    & e^{\mathit{j} \theta_t^2} & &  \\
    &  & \ddots & \\
    & &  & e^{\mathit{j} \theta_t^L}
  \end{bmatrix}.
\end{equation}

The signal received at the malicious detector comprises three components: (i) the target echo induced by the probing signal transmitted from the malicious detector, (ii) the target echo generated by the probing signal emitted from the ISAC transmitter and forwarded to the target via the STAR-RIS, and (iii) the ISAC communication signal transmitted via the STAR-RIS. Note that, from the malicious detector's perspective, the last component acts as interference and reduces its target detection capability. Based on this, the received signal at malicious detector when the target exists is given by 

\vspace{-0.5cm}
\begin{equation}
\label{tab:s1}
y_{s|d} = \eta \sqrt{p_d}|d|^2x_d(t)+\eta \sqrt{p_s} d\mathbf{h}_{\text{s|eff}}\mathbf{x}(t)+\sqrt{p_c}\mathbf{i}_{\text{eff}}\mathbf{x}(t)+n_s(t), 
\end{equation}

\noindent where $\eta$ is the target radar cross section with $\zeta^2 = \mathbb{E}(|\eta|^2)$, $p_d$ is the malicious detector transmission power, $p_s$ is the ISAC transmitter's probing signal power, and $d$ denotes the line-of-sight (LoS) channel between the malicious detector and the target. The value of $d$ is given by
\vspace{-0.5cm}

\begin{equation}
\label{tab:dd}
d = \alpha_d \exp\bigg(j\pi\cos(\beta_d)\sin(\phi_d)+\sin(\beta_d)\sin(\phi_d)\bigg),
\end{equation}

\noindent where $\alpha_d$, $\beta_d$, and $\phi_d$ represent the malicious detector-target's path gain, elevation angle, and azimuth angle, respectively.  $x_d(t)$ is the malicious detector sensing signal with $|x_d(t)|^2=1$, $x(t)$ is the ISAC transmitter's signal with $|x(t)|^2=1$, and $n_s(t)$ is the additive white Gaussian noise (AWGN). Accordingly, the received sensing SINR at the  malicious detector is given by

\begin{equation}
\label{tab:SINRs1}
\gamma_{s|d} = \frac{\zeta^2p_d|d|^4+\zeta^2p_s|d|^2\|\mathbf{h}_{\text{s|eff}}\|^2}{\sigma_d^2+p_c\|\mathbf{i}_{\text{eff}}\|^2}. 
\end{equation}

Similarly, when the target exists, the signal received at the ISAC receiver comprises four components: (i) the target echo generated by the probing signal emitted from the ISAC transmitter, (ii) the target echo induced by the probing signal transmitted from the malicious detector, (iii) the interference from the malicious detector's sensing signal, and (iv) the interference from the jamming signal \footnote{Similar to the adversarial nodes coordination, the ISAC communication signal is nulled at the ISAC receiver.}. Therefore, the received signal at the ISAC receiver is given by

\begin{equation}
\label{tab:s2}
\begin{aligned}
\mathbf{y}_{s|r} &= \eta \sqrt{p_s}r\mathbf{h}_{\text{s|eff}}\mathbf{x}(t)+\eta \sqrt{p_d} r d x_d(t) \\
&+ \sqrt{p_d}h_{r,d}x_d(t)+ \sqrt{p_j}\mathbf{h}_{\text{R-J|eff}}\mathbf{x}_j(t)+n_d(t), 
\end{aligned}
\end{equation}

\noindent where $h_{r,d}$ is the direct channel between the ISAC receiver and the malicious detector and $r$ is the LoS channel between the ISAC receiver and the target. Similar to (\ref{tab:dd}),  the value of $r$ is given by

\begin{equation}
\label{tab:rr}
r = \alpha_r \exp\bigg(j\pi\cos(\beta_r)\sin(\phi_r)+\sin(\beta_r)\sin(\phi_r)\bigg),
\end{equation}

\noindent where $\alpha_r$, $\beta_r$, and $\phi_r$ represent the ISAC receiver-target's path gain, elevation angle, and azimuth angle, respectively. Afterward, we can derive the received SINR at the ISAC receiver as follows

\begin{equation}
\label{tab:SINRs2}
\gamma_{s|r} = \frac{p_s\zeta^2\|\mathbf{r}\|^2\|\mathbf{h}_{\text{s|eff}}\|^2+ \zeta^2p_d|d|^2\|\mathbf{r}\|^2}{\sigma_d^2+p_d\|h_{r,d}\|^2+p_j\|\mathbf{h}_{\text{R-J|eff}}\|^2}, 
\end{equation}

On the other hand, when the target doesn't exist, the received signals at the malicious detector and the ISAC receiver are given respectively by

\begin{equation}
\label{tab:s3}
    \begin{aligned}
     \tilde{y}_{s|d} &= \sqrt{p_c}\mathbf{i}_{\text{eff}}\mathbf{x}(t)+n_s(t), \\
    \tilde{\mathbf{y}}_{s|r} &= \sqrt{p_d}\mathbf{h}_{r,d}x_d(t)+\sqrt{p_j}\mathbf{H}_{\text{R-J|eff}}\mathbf{x}_j(t)+n_d(t).
    \end{aligned}
\end{equation}

The next step is to derive the detection probability at the malicious detector and the ISAC receiver. Based on (\ref{tab:s1}) -(\ref{tab:s3}), it can be shown that the probabilities of target detection at the malicious detector and the ISAC receiver are expressed, respectively, as follows\cite{refHide}

\begin{equation}
\label{tab:pd}
\begin{aligned}
   & \mathcal{P}_d = \left(1+\gamma_{s|d}\right)^{-\gamma_{s|d}^{-1}}, \\
   & \mathcal{P}_{d|ISAC} = \left(1+\gamma_{s|r}\right)^{-\gamma_{s|r}^{-1}}
\end{aligned}
\end{equation}

\section{Problem Formulation}\label{tab:probform}

In this paper, we utilize the STAR-RIS to mitigate the impact of jamming on communication quality and conceal the target from adversarial detection. To this end, the STAR-RIS reflection coefficients are optimized to suppress jamming interference, while the transmission coefficients are configured to minimize the malicious detector's detection probability. Consequently, the joint STAR-RIS optimization problem is formulated as follows:

\begin{equation}
\label{tab:p1}
\begin{aligned}
& \underset{\boldsymbol{\Psi}_t, \boldsymbol{\Psi}_r}{\text{\textbf{$\min$}},}
& & f(\boldsymbol{\Psi}_t, \boldsymbol{\Psi}_r)= w_1\|\mathbf{J}_{\text{eff}}\|^2+w_2\mathcal{P}_d,\\
& \text{subject to}
& & (a) \|\mathbf{H}_{\text{eff}}\|^2 \geq G_{\text{th}},\\
& & & (b) \mathcal{P}_{d|ISAC} \geq \mathcal{P}_{d_{\text{min}}},\\
& & & (c) |e^{\mathit{j} \theta_{t}^l}| = 1 \text{ 
 } \forall l\in\{1,2,\cdots L\},\\
& & & (d) |e^{\mathit{j} \theta_{r}^{l}}| = 1 \text{  } \forall l\in\{1,2,\cdots L\},\\
\end{aligned}
\tag{\textbf{$\mathbf{P}$}}
\end{equation}

\noindent where constraint (a) ensures that the  communication channel gain meets a predefined threshold, $G_{\text{th}}$, such that $\|\mathbf{H}_{\text{eff}}\|^2 \geq G_{\text{th}}$. Constraint (b) enforces a minimum sensing requirement by ensuring that the ISAC receiver's detection probability remains above $\mathcal{P}_{d_{\text{min}}}$. Finally, constraints (c) and (d) ensure that the STAR-RIS satisfies the passivity requirements. 

The problem is (\ref{tab:p1}) is a non-convex optimization problem. However, it is fully separable into the reflection and transmission subspaces. Accordingly, the problem in (\ref{tab:p1}) can be decomposed into two independent subproblems optimizing the STAR-RIS transmission and reflection phase shifts, respectively. 
\vspace{-0.5cm}

\subsection{STAR-RIS Transmission-Aided Target Concealment.}

The first subproblem aims to conceal the target by minimizing the detection probability at the malicious detector through the optimization of the STAR-RIS transmission phase shifts as follows.

\begin{equation}
\label{tab:p11}
\begin{aligned}
& \underset{\boldsymbol{\Psi}_t}{\text{\textbf{$\min$}},}
& & f(\boldsymbol{\Psi}_t)= \mathcal{P}_d,\\
& \text{subject to}
& & (b) \mathcal{P}_{d|ISAC} \geq \mathcal{P}_{d_{\text{min}}},\\
& & &  (c) |e^{\mathit{j} \theta_{t}^{l}}| = 1 \text{  } \forall l\in\{1,2,\cdots L\}.\\
\end{aligned}
\tag{\textbf{$\mathbf{P1}$}}
\end{equation}

\noindent From (\ref{tab:pd}), it can be seen that both $f(\boldsymbol{\Psi}_t) = \mathcal{P}_d$ and $\mathcal{P}_{d|ISAC}$ are both monotonically increasing with $\gamma_{s|d}$ and $\gamma_{s|r}$, respectively. Accordingly, the problem in (\ref{tab:p11}) can be reformulated as follows
\vspace{-0.5cm}
\begin{equation}
\label{tab:p111}
\begin{aligned}
& \underset{\boldsymbol{\Psi}_t}{\text{\textbf{$\min$}},}
& & \gamma_{s|d},\\
& \text{subject to}
& & (b) \gamma_{s|r} \geq \gamma_{min},\\
& & &  (c) |e^{\mathit{j} \theta_{r}^{l}}| = 1 \text{  } \forall l\in\{1,2,\cdots L\}.\\
\end{aligned}
\tag{\textbf{$\mathbf{P1.1}$}}
\end{equation}

\noindent The problem in (\ref{tab:p11}) is still a non-convex optimization problem, making it challenging to find its global optimal solution. Accordingly, we employ Fractional Programming (FP) and the Dinkelbach method to address the fractional form of the SINR objective. Specifically, we reformulate the optimization problem in terms of the matrix variable $\mathbf{Y}_t = \psi_t\psi_t^H$, where $\psi_t = [e^{\mathit{j} \theta_{t}^{1}}, e^{\mathit{j} \theta_{t}^{2}}, ..., e^{\mathit{j} \theta_{t}^{L}}]^H$. Since $|e^{\mathit{j} \theta_{r}^{l}}| = 1 \text{ }\forall l \in \{1,2, ...L\}$, it follows that $\mathbf{Y}_t$ is a positive semidefinite matrix, i.e., $\mathbf{Y_t\succeq0}$, with $\text{rank}(\mathbf{Y}_t)=1$. Furthermore, we express the effective channel vectors  the jammer to the ISAC receiver, $\mathbf{h}_{\text{R-J|eff}} = \psi_t^H \mathbf{A}_r = \psi_t^H \text{diag}( \mathbf{f}_r)\mathbf{G}_j$, from the ISAC transmitter to the target, $\mathbf{h}_{\text{s|eff}} = \psi_t^H \mathbf{A}_s = \psi_t^H\text{diag}( \mathbf{f}_t)\mathbf{G}_b$, and from the ISAC transmitter to the malicious detector, $\mathbf{i}_{\text{eff}}  = \psi_t^H \mathbf{A}_m = \psi_t^H\text{diag}( \mathbf{f}_m)\mathbf{G}_b$. Accordingly, the reformulated problem is given by

\footnotesize
\begin{equation}
\label{tab:p12}
\begin{aligned}
& \underset{\mathbf{Y}_t}{\text{\textbf{$\min$}},}
& &\bigg(\zeta^2p_d|d|^4+\zeta^2p_c|d|^2\text{Tr}(\mathbf{A}_s\mathbf{Y}_t)\bigg) -\lambda_s \bigg(\sigma_d^2+p_c \text{Tr}(\mathbf{A}_m\mathbf{Y}_t)\bigg). \\
& \text{subject to}
& & (b) \frac{p_c\zeta^2\|\mathbf{r}\|^2\text{Tr}(\mathbf{A}_s\mathbf{Y}_t)+ \zeta^2p_d|d|^2\|\mathbf{r}\|^2}{\sigma_d^2+p_d\|\mathbf{h}_{r,d}\|^2+p_j\text{Tr}(\mathbf{A}_r\mathbf{Y}_t)}\geq \gamma_{min},\\
& & &  (c.1)\text{rank}(\mathbf{Y}_t)= 1 ,\\
& & &  (c.2)\mathbf{Y}_t\succeq 0 ,\\
\end{aligned}
\tag{\textbf{$\mathbf{P1.2}$}}
\end{equation}
\normalsize

\noindent Finally, to solve the non-convex problem in (\ref{tab:p12}), we relax the rank-one constraint. Consequently,The problem becomes a standard SDP problem which can be solved by CVX \cite{CVX}. Afterwards, Gaussian randomization is utilized to retrieve the rank-one solution from the relaxed martrix solution.   

\subsection{STAR-RIS Reflection-Aided Jamming Mitigation}

\noindent The second subproblem aims to mitigate the jamming interference through the optimization of the STAR-RIS reflection phase shifts as follows.

\begin{equation}
\label{tab:p2}
\begin{aligned}
& \underset{ \boldsymbol{\Psi}_r}{\text{\textbf{$\min$}},}
& & f( \boldsymbol{\Psi}_r)= \|\mathbf{J}_{\text{eff}}\|^2,\\
& \text{subject to}
& & (a)  \|\mathbf{H}_{\text{eff}}\|^2 \geq G_{\text{th}},\\
& & & (d) |e^{\mathit{j} \theta_{r}^{l}}| = 1 \text{  } \forall l\in\{1,2,\cdots L\}.\\
\end{aligned}
\tag{\textbf{$\mathbf{P2}$}}
\end{equation}
 \noindent 

\noindent The problem in (\ref{tab:p2}) is a non-convex optimization problem due to the presence of constraint $(d)$. However, by adopting a penalty approach to account for the violation of constraint $(a)$, the problem can be reformulated as an unconstrained optimization problem on the surface of the  complex circle manifold. This is then solved using the Riemannian Polak-Reib\'{e}re conjugate gradient algorithm. The reformulated STAR-RIS reflection optimization problem is given by 

\begin{equation}
\label{tab:p21}
\begin{aligned}
& \underset{ \boldsymbol{\Psi}_r}{\text{\textbf{$\min$}},}
& & f_2( \boldsymbol{\Psi}_r)= \|\mathbf{J}_{\text{eff}}\|^2 + \lambda_r (G_{\text{th}}-\|\mathbf{H}_{\text{eff}}\|^2),\\
& \text{subject to}
& & (d) |e^{\mathit{j} \theta_{r}^{l}}| = 1 \text{  } \forall l\in\{1,2,\cdots L\}.\\
\end{aligned}
\tag{\textbf{$\mathbf{P2.1}$}}
\end{equation}
 \noindent The  Riemannian Polak-Reib\'{e}re conjugate gradient algorithm begins by calculating the Euclidean gradient of the objective function, $f_2(\boldsymbol{\Psi}_r)$ with respect to the vector  $\boldsymbol{\psi}_r = [\psi_{1r}=e^{\mathit{j} \theta_{r}^{1}}, \psi_{2r} = e^{\mathit{j} \theta_{r}^{2}}, ..., \psi_{Lr}=e^{\mathit{j} \theta_{r}^{L}}]^T$ as follows.

\footnotesize
\begin{equation}
\label{tab:egrad}
\begin{aligned}
& \nabla_{\psi_{rl}} f_2(\boldsymbol{\Psi}_r) =
\text{Tr}\left(\left(\nabla_{\mathbf{J}_{\text{eff}}}f_2(\boldsymbol{\Psi}_r)\right)^\dagger \times \frac{\delta J_{\text{eff}}}{\delta \psi_{rl}} \right) + \text{Tr}\left(\left(\nabla_{\mathbf{J}_{\text{eff}}}f_2(\boldsymbol{\Psi}_r)\right) \times \frac{\delta J_{\text{eff}}}{\delta \psi_{rl}}^\dagger \right) \\
     & + \lambda_r \text{Tr}\left(\left(\nabla_{\mathbf{H}_{\text{eff}}}f_2(\boldsymbol{\Psi}_r)\right)^\dagger \times \frac{\delta H_{\text{eff}}}{\delta \psi_{rl}} \right) + \text{Tr}\left(\left(\nabla_{\mathbf{H}_{\text{eff}}}f_2(\boldsymbol{\Psi}_r)\right) \times \frac{\delta H_{\text{eff}}}{\delta \psi_{rl}}^\dagger \right),
\end{aligned}
\end{equation}
\normalsize

\noindent where $\nabla_{\mathbf{J}_{\text{eff}}} = 2J_{\text{eff}}$, $\nabla_{\mathbf{H}_{\text{eff}}} = -2H_{\text{eff}}$,  $\frac{\delta \mathbf{J}_{\text{eff}}}{\delta \psi_{rl}} =  \left[\mathbf{F}_c\right]_{:, \ell} \otimes \left[\mathbf{G}_{j}\right]_{\ell, :}$, and $\frac{\delta \mathbf{H}_{\text{eff}}}{\delta \psi_{rl}} =  \left[\mathbf{F}_c\right]_{:, \ell} \otimes \left[\mathbf{G}_{b}\right]_{\ell, :}$. Afterward, the Riemannian gradient is derived by orthogonally projecting the Euclidean gradient onto the tangent space of the complex circle manifold. This gradient is then utilized determine the search direction and update the variables via the Riemannian Polak-Reib\'{e}re conjugate gradient \cite{absil}. 


\begin{figure}[t]
    \centering
    \includegraphics[width=9cm,height=7.9cm]{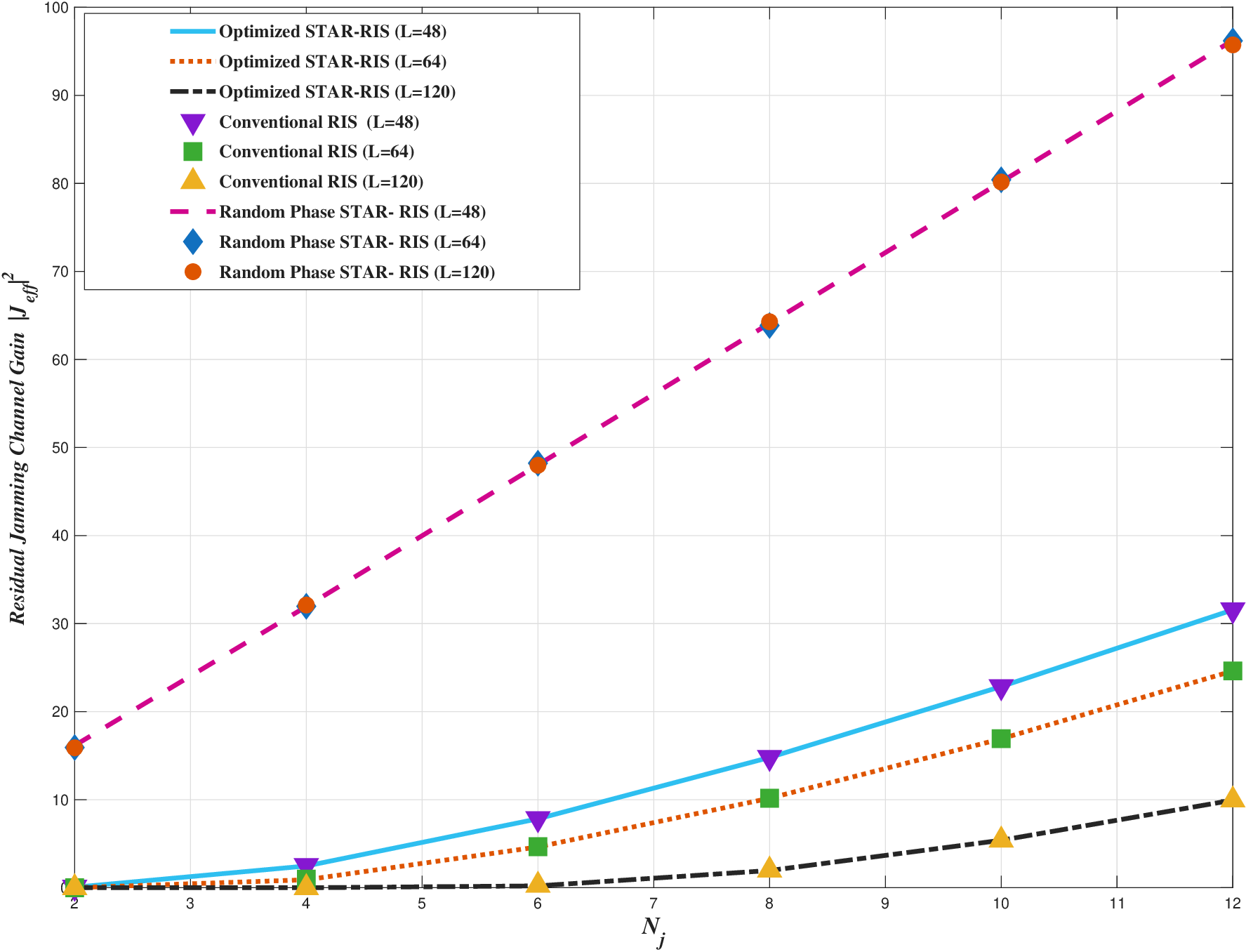} 
    \caption{Residual jamming Vs $N_j$.}
    \label{tab:resJamNj}
  \end{figure}

\section{Numerical Analysis}\label{tab:numAna}

In this section, we evaluate the proposed SAFE-ISAC framework through extensive numerical simulations. Its performance is assessed by comparing the resulting malicious SINR and residual jamming gain against two benchmarks: a random-phase STAR-RIS-assisted model and a conventional reflective RIS-assisted model. Unless otherwise specified, the simulation parameters are set to $N=16$, $N_j =4$, $K=4$, $p_c = 6$dBW, $p_s=8$ dBW, $p_j=p_d = 30$ dBm, and the AWGN noise variance $\sigma^2 = -120$ dBm. 

First, we evaluate the residual jamming channel gain and the malicious detector SINR as a function of the number of jamming antennas, $N_j$. As shown in Fig. \ref{tab:resJamNj}, the residual jamming channel gain increases with $N_j$ due to enhanced antenna gain available to the jammer. Nevertheless, both STAR-RIS and conventional reflective RIS effectively reduce the jamming interference. While one might initially infer that a conventional RIS is sufficient, its limitations become apparent when examining the malicious detector's SINR.

\begin{figure}[t]
    \centering
    \includegraphics[width=9cm,height=7.9cm]{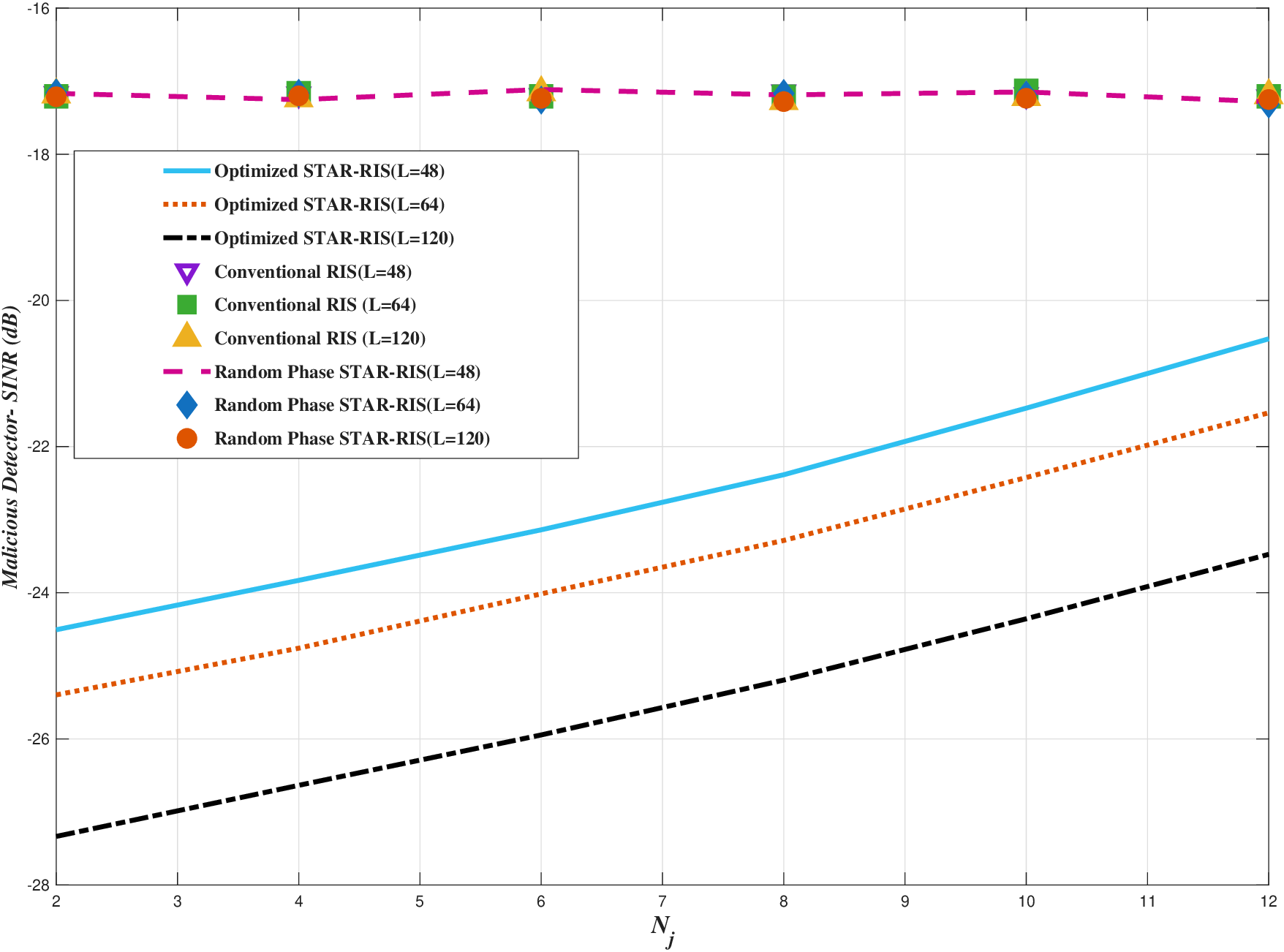} 
    \caption{Malicious detection SINR Vs $N_j$}
    \label{tab:sinrJamNj}
  \end{figure}

 As illustrated in Fig. \ref{tab:sinrJamNj}, the conventional RIS fails to provide sufficient protection against the malicious detector, achieving nearly the same SINR as the random-phase STAR-RIS. This indicates that conventional RIS's limited degrees of freedom are exhausted by satisfying QoS requirements and mitigating jamming, leaving no room for target concealment. In contrast, due to its superior environmental control, STAR-RIS simultaneously satisfies QoS constraints, suppresses jamming, and significantly reduces the malicious detector's SINR. 

Another interesting observation is that increasing the number of jamming antennas unexpectedly enhances the malicious detector's SINR. This occurs because the jammer-detector coordination ensures that the higher jamming power only degrades the target's echo received at the ISAC receiver $\gamma_{s|r}$. To maintain  $\gamma_{s|r} \geq \gamma_{min}$ defined in (\ref{tab:p111}), the feasible region of the optimization problem diminshes. Consequently, the STAR-RIS is forced to prioritize sensing reliability, leaving fewer degrees of freedom to suppress the SINR at the malicious detector.

\begin{figure} [t]
    \centering
    \includegraphics[width=9cm,height=7.9cm]{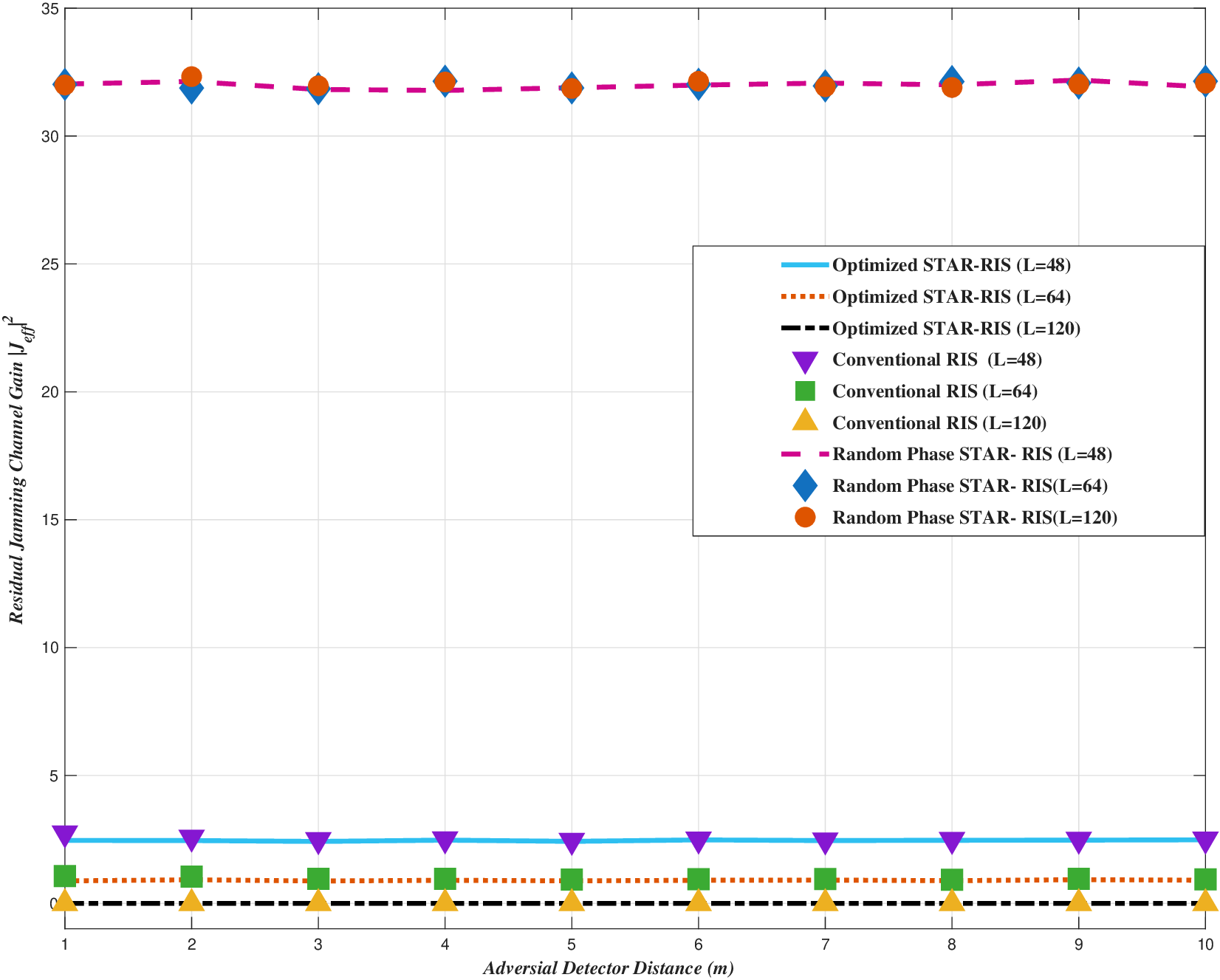} 
    \caption{Residual jamming Vs malicious detector's distance.}
    \label{tab:resJamdmal}
  \end{figure}

\indent Next, the system performance is further evaluated with respect to the malicious detector's distance from the target. As illustrated in Figs. \ref{tab:resJamdmal} and \ref{tab:sinrJamdmal}, varying the malicious detector-to-target distance impacts its SINR but does not influence jamming mitigation. Specifically, Fig. \ref{tab:resJamdmal} shows that the residual jamming channel gain remains invariant to changes in the detector's location. While the conventional RIS achieves jamming mitigation comparable to the proposed STAR-RIS, the advantages of the SAFE-ISAC framework are particularly evident in the malicious detector's SINR. As shown in Fig. \ref{tab:sinrJamdmal}, the reflective-only and random-phase benchmarks fail to provide adequate protection. In contrast, the proposed STAR-RIS suppresses the malicious SINR by approximately 8 dB for $L=48$ and 10 dB for $L=120$. This demonstrates that increasing the number of STAR-RIS elements, $L$, significantly enhances the concealment capabilitity, providing a more robust defense against malicious detection.

\section{Conclusion}\label{tab:conc}

This paper investigated a STAR-RIS-aided bistatic ISAC system designed to operate under a coordinated adversarial attack. We proposed a joint optimization framework, SAFE-ISAC, to simultaneously suppress active communication jamming within its reflection subspace and minimize target detectability by a malicious detector within its transmission subspace. Numerical results demonstrated that while conventional reflective-only RIS is effective for jamming suppression, it lacks the degrees of freedom required to simultaneously protect the sensing target. On the other hand, the proposed STAR-RIS architecture can effectively  maintain the required communication and sensing QoS constraints. Furthermore, the results demonstrate its significant capability in reducing both received jamming power and the malicious detector's SINR compared to conventional reflecting RIS and random-phase RIS benchmarks.

\bibliographystyle{IEEEtran}

\begin{thebibliography}{999}
\bibitem{6G} CH. Wang, X. You, X. Gao, X. Zhu, Z. Li, C. Zhang, H. Wang et al. "On the road to 6G: Visions, requirements, key technologies, and testbeds." \emph{IEEE Communications Surveys \& Tutorials,} vol. 25, no. 2, pp. 905-974, Feb. 2023. 

\bibitem{noma1} A. Ahmed, X. Wang, A. Hawbani, W. Yuan, H. Tabassum, Y. Liu et al. "Unveiling the potential of NOMA: A journey to next-generation multiple access." \emph{IEEE Communications Surveys \& Tutorials}, vol. 27, no. 5, pp. 3099-3164, Oct. 2025. 

\bibitem{noma2} Y. Liu, S. Zhang, X. Mu, Z. Ding, R. Schober, N. Al-Dhahir, E. Hossain, and X. Shen. "Evolution of NOMA toward next generation multiple access (NGMA) for 6G." \emph{IEEE Journal on Selected Areas in Communications}, vol. 40, no. 4, pp. 1037-1071, Apr. 2022. 

\bibitem{oam1} Z. Wang, M. El-Hajjar, and LL Yang. "Orbital angular momentum for wireless communications: Key performance indicators and performance comparison." \emph{IEEE Access}, pp. 80889- 80913, May 2025.

\bibitem{oam2} H. Zhang, Z. Cao, H. Xie, and H. Jin. "Orbital angular momentum (OAM) in wireless communication: Applications and challenges towards 6G." In \emph{14th International Conference on Information and Communication Technology Convergence (ICTC)}, Jeju Island, Republic of Korea, Oct. 2023. 

\begin{figure}[t]
    \centering
    \includegraphics[width=9cm,height=7.9cm]{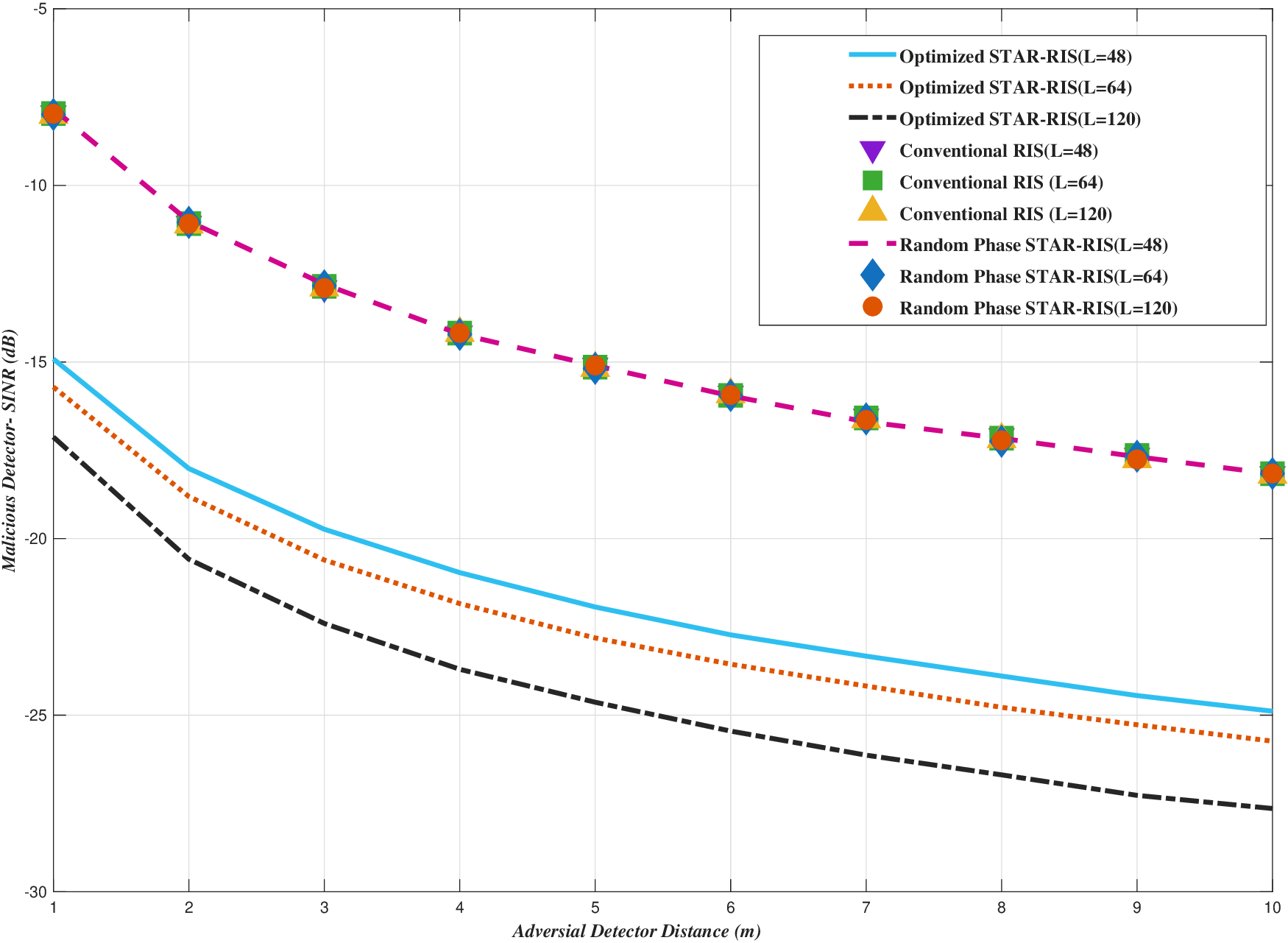} 
    \caption{Malicious detection SINR Vs malicious detector's distance}
    \label{tab:sinrJamdmal}
  \end{figure}

\bibitem{fd1} B. Smida, A. Sabharwal, G. Fodor, G.C. Alexandropoulos, H.A. Suraweera, and CB. Chae. "Full-duplex wireless for 6G: Progress brings new opportunities and challenges." \emph{IEEE Journal on Selected Areas in Communications}, vol. 41, no. 9, pp. 2729-2750, Jun. 2023.

\bibitem{fd2} R. Sultan, and A. Shamseldeen. "Uplink-downlink cochannel interference cancellation in RIS-Aided full-duplex networks." \emph{IEEE Systems Journal}, vol. 18, no. 2, pp. 1220-1223, Jun. 2024. 

\bibitem{fd3} R. Sultan, and A. Shamseldeen. "STAR-RIS-aided full-duplex communication for massive MIMO IoT systems." In \emph{IEEE 15th International Conference on Computational Intelligence and Communication Networks (CICN)}, Bangkok, Thailand, Dec. 2023. 


\bibitem{isac1} F. Liu, Y. Cui, C. Masouros, J. Xu, T.X. Han, Y. C. Eldar, and S. Buzzi. "Integrated sensing and communications: Toward dual-functional wireless networks for 6G and beyond." \emph{IEEE journal on selected areas in communications}, vol. 40, no. 6, pp. 1728-1767, Mar. 2022. 

\bibitem{isac2} Z. Wei, H. Qu, Y. Wang, X. Yuan, H. Wu, Y. Du, K. Han, N. Zhang, and Z. Feng. "Integrated sensing and communication signals toward 5G-A and 6G: A survey." \emph{IEEE Internet of Things Journal}, vol. 10, no. 13, pp. 11068-11092, Jul. 2023. 

\bibitem{isac3} Y. Liu, M. Li, Y. Han, and L. Ong. "Fundamental Limits of Integrated Sensing and Communication over Interference Channels." \emph{IEEE Journal on Selected Areas in Information Theory}, vol. 7, pp. 1-15, Jan. 2026. 

\bibitem{ris1} Y. Zhou, W. Xu, W. Xiang, and Y. Wu. "Multi-Functional RIS-Aided ISAC Systems: Beamforming Design and Performance Analysis." \emph{IEEE Transactions on Vehicular Technology}, Early Access, Feb. 2026. 

\bibitem{ris2} J. Ye, Jianglin, J. Dai, C. Pan, K. Wang, and J. Li. "Joint active and passive beamforming design for secure RIS-aided ISAC system." \emph{IEEE Wireless Communications Letters} vol. 14, no. 3, pp. 916-920, Mar. 2025

\bibitem{ris3} W. Ma, P. Zhang, J. Ye, R. Guan, XP Li, and L. Huang. "Joint Antenna Selection and Beamforming Design for Active RIS-Aided ISAC Systems." \emph{IEEE Internet of Things Journal}, vol. 12, no. 14, pp. 26500-26513, Jul. 2025.


\bibitem{refHide} A. Magbool, V. Kumar, M. D. Renzo, and M. F. Flanagan. "Hiding in plain sight: Ris-aided target obfuscation in isac." arXiv preprint arXiv:2503.05418 (2025).


\bibitem{sec1} F. Yang, C.  Xing, H. Wei, M. Jo, N. Deng, N. Zhao, and D. Niyato. "Intelligent Covert ISAC via RIS: A Reinforcement Learning Approach." \emph{IEEE Transactions on Wireless Communications}, vol. 25, pp. 13107-13120, Mar. 2026. 

\bibitem{sec2} L. Guo, J. Jia, X. Mu, Y. Liu, J. Chen, and X. Wang. "Joint secure and covert communications for active STAR-RIS assisted ISAC systems." \emph{IEEE Transactions on Wireless Communications}, vol. 24, no. 9, pp. 7501-7516, Sep. 2025.



\bibitem{refpower} S. Lin, X. Li, H. Xu, and S. Jin. "Energy efficiency optimization in RIS-assisted communication systems with RIS power budget," \emph{IEEE Transactions on Vehicular Technology}, Early Access, Aug. 2025.

\bibitem{ref222} Y. Sun, K. An, J. Luo, Y. Zhu, G. Zheng, and S. Chatzinotas. "Intelligent reflecting surface enhanced secure transmission against both jamming and eavesdropping attacks." \emph{IEEE Transactions on Vehicular Technology}, vol.70, no. 10, pp. 11017-11022, Oct. 2021. 



\bibitem{CVX} Michael Grant and Stephen Boyd. CVX: Matlab software for disciplined convex programming, version 2.0 beta. https://cvxr.com/cvx, Sep. 2013.

\bibitem{absil} P-A. Absil, R. Mahony, and R. Sepulchre. Optimization algorithms on matrix manifolds. Princeton University Press, 2008.

\end{thebibliography}

\end{document}